\documentclass[twocolumn,preprintnumbers,amsmath,amssymb]{revtex4}

\usepackage{graphicx}
\usepackage{dcolumn}
\usepackage{bm}

\newcommand{\be}{\begin{eqnarray}}
\newcommand{\ee}{\end{eqnarray}}

\begin{document}

\title{\ The Dark Side of Strongly Coupled Theories}

\author{Chris Kouvaris}

\affiliation{The Niels Bohr Institute, Blegdamsvej 17, DK-2100 Copenhagen,
Denmark}

\date{\today}

\begin{abstract}
We investigate the constraints of dark matter search experiments on the different candidates emerging from the minimal quasi-conformal strong coupling theory with fermions in the adjoint representation. For one candidate, the current limits of CDMS exclude a tiny window of masses around 120 GeV. We also investigate under what circumstances the newly proposed candidate composed of a $-2$ negatively charged particle and a $^4He^{+2}$ can explain the discrepancy between the results of the CDMS and DAMA experiments. We found that this type of dark matter should give negative results in CDMS, while it can trigger the detectors of DAMA, if a condition between the mass and the binding energy of the $-2$ particle with the nucleus of the detector is satisfied.
\end{abstract}

\maketitle


\section{Introduction}
New technicolor theories with techniquarks transforming under higher dimensional representations
have become lately the focus of interest. They offer a natural dynamical way of breaking the electroweak
symmetry without violating the experimental constraints~\cite{Sannino:2004qp,Hong:2004td,Dietrich:2005jn,Dietrich:2006cm}. Such theories can enjoy a walking behavior, thus
solving the problem of giving mass to the heaviest Standard Model particles, with a very small number of colors
and flavors. In particular, the minimal model can be quasi-conformal with only two techniquarks and two colors,
as long as the techniquarks transform under the adjoint representation of $SU(2)$. The particle content of this
 minimal model apart from the two techniquarks $U$ up and $D$ down, includes an extra family of heavy leptons
 $\nu'$ and $\zeta$, in order to cancel the Witten global anomaly. The global symmetry between the techniquarks is
 enhanced from the usual $SU(2)_L \times SU(2)_R$ to an $SU(4)$ due to the fact that techniquarks transform under the
 adjoint representation of the gauge group. The chiral condensate of the theory is invariant under an $SO(4)$ symmetry that
 includes the $SU(2)_V$ as a subgroup. The effective theory of the minimal model was studied extensively in~\cite{Gudnason:2006ug,Foadi:2007ue}. Holographic~\cite{Dietrich:2008ni} and lattice methods have been also used as tools in the study of the model~
\cite{Catterall:2007yx,DelDebbio:2008wb,DelDebbio:2008zf,Sannino:2008ha,Catterall:2008qk}.

Although simple and minimal in terms of particle content, the minimal model can
 incorporate several dark matter scenarios. It was pointed out that for a specific choice of the hypercharge for the technicolor
 particles, that does not induce any gauge anomalies, one of the techniquarks can become electrically neutral~\cite{Gudnason:2006yj}. We can choose
 $D$ to be the one, although it is a matter of convention, and the results can be easily extracted if it is the $U$ instead of $D$.
 So far, four different dark matter scenarios have been investigated based on the minimal technicolor model. In the first one~\cite{Gudnason:2006yj},
 the dark matter particle is the $DD$ Nambu-Goldstone boson of the theory, that is electrically neutral. If it is also the case
 that this particle is the lightest technibaryon, it should be stable provided there are no technibaryon violating processes. Although $DD$ is a Nambu-Goldstone boson, it is a diquark carrying technibaryon number, rather than a
 meson of the pion type, that is made of a quark-antiquark pair. The electroweak sector posses technibaryon violating processes
 through sphalerons. As in the Standard Model where the baryon number $B$ and lepton number $L$ are violated separately through anomalies, also in this case the technibaryon number $TB$ as well as the number of the ``new lepton family'' $L'$ are violated due to sphaleron processes. However, such processes are exponentially suppressed once the temperature drops below the electroweak scale. It was shown in~\cite{Gudnason:2006yj}, that for a mass of the order of TeV, $DD$ can have the proper relic density to
 account for the dark matter in the Universe. However, the elastic cross section of $DD$ scattering off nuclei targets in dark matter search experiments like CDMS, excludes $DD$ as a primary WIMP (weakly interacting massive particle). The current
 exposure of the $Ge$ detectors in CDMS is 121 kg.days~\cite{Ahmed:2008eu}. The negative search results put tight constraints on the elastic
 cross section of WIMPs, provided that the dark matter density around the Earth is roughly 0.3 GeV$/\text{cm}^3$. This excludes
 $DD$ as a WIMP by several orders of magnitude, unless $DD$ consists a tiny component of dark matter or its mass is unrealistically high.

 The exclusion of the remaining three scenarios is more subtle. The first one is the colorless and electrically neutral bound state $DG$ between a $D$ and a technigluon $G$, studied in~\cite{Kouvaris:2007iq}. A second possibility is the $\nu'$ particle (for a different hypercharge assignment, namely the one that makes $\nu'$ neutral)~\cite{Kainulainen:2006wq}, and the third option is to have a state bound by electromagnetic Coulomb forces between a $-2$ electrically charged  $\bar{U}\bar{U}$ or $\zeta$ and a $^4He^{+2}$~\cite{Khlopov:2007ic,Khlopov:2008ty}. In this paper we derive the required exposure in kg.days in the CDMS $Ge$ detectors, in order to detect
$DG$ or $\nu'$ taking into account the new CDMS results, effects from the motion of the Earth and a more precise formula for the elastic cross section of $DG$ scattered off nuclei targets valid also in the regime where the mass of $DG$ is comparable with the one of the $W$ boson. In addition, we provide the calculation of relic density of techni-O-helium (the $\bar{U}\bar{U}$ or $\zeta$ and $^4He^{+2}$ bound states) for the case of a first order phase transition for the electroweak symmetry breaking. Finally, we speculate on what conditions such a candidate can explain the discrepancy between the findings of the CDMS~\cite{Ahmed:2008eu} and DAMA~\cite{Bernabei:2008yi} experiments as it was suggested in~\cite{Khlopov:2008ty,Khlopov:2008ki}.

\section{The case of DG}

In this particular case, and for the hypercharge anomalous free choice, where the electric charge of $D$ is zero, $D$ and technigluons can form bound colorless states. If $DG$ is the lightest techniparticle, it can account for the dark matter density of the Universe, avoiding the same time detection from CDMS for almost any mass of interest. This can happen in the case where $D_LG$ has a Majorana mass, and a seesaw mechanism takes place. The lightest particle $N_2$, couples to the electroweak sector through its left-handed part, and the strength of the coupling is controlled by the mixing angle between left and right-handed $D$ techniquarks $\theta$, defined in~\cite{Kouvaris:2007iq}. It is worth mentioning that the technibaryon number is not preserved due to the Majorana mass, however there is a $Z_2$ symmetry instead, similar to the R-parity of the neutralino. Such a particle might be subject to indirect signatures as proposed in~\cite{Kouvaris:2007ay}. Two $N_2$ can co-annihilate through a $Z$ boson mediation to Standard Model particle-antiparticle pairs. As it was argued in~\cite{Kouvaris:2007iq}, $N_2$ can account for the dark matter density with the proper adjustment of the mixing angle $\theta$, if it is heavier than 23 GeV. The annihilation cross section is dominated by the annihilation channels to light quark-antiquark and lepton-antilepton pairs for small masses (smaller than roughly 80 GeV) and by $W^+$-$W^-$ production as soon as this channel opens up at 80 GeV. The cross section was calculated in both regimes in~\cite{Kouvaris:2007iq}, although in the large mass regime, the cross section was calculated only at the limit where $m>>M_W$, where $m$ is the mass of $N_2$. Here, we calculated the annihilation cross section to $W^+$-$W^-$ production valid not only for $m>>M_W$, but also at the onset of the channel where $m\sim M_W$,
\be \sigma = \frac{G_F^2 \beta_W v^2}{12 \pi} D_Z^2 (s-4M_W^2)(s-2M_W^2)^2\sin^4\theta, \label{s}\ee
where $D_Z^2=1/((s-M_Z^2)^2+ \Gamma_Z^2 M_Z^2)$, $\beta_W$ and $v$ are the velocities of the $W$ and $N_2$ respectively at the center of mass reference system, $s \simeq 4m^2$, and $\Gamma_Z$ and $M_Z$ are the width and the mass of $Z$. We can now calculate the relic density of $N_2$ as in~\cite{Kouvaris:2007iq}. We also include the annihilation channels to pairs of light fermions. Fig.~1 shows the value of the mixing angle $\sin\theta$ as a function of $m$, such that $N_2$ has a relic density equal to the dark matter density of the Universe. For comparison, we have also plotted the value of $\sin\theta$ acquired in~\cite{Kouvaris:2007iq}. The figure shows that the effect of $M_W$ changes the result only for $m$ between 80 and 250 GeV. At the peak of the curve, the difference is about 12$\%$, and above 250 GeV, the two curves effectively coincide.

\begin{figure}[t]
\begin{center}
\includegraphics[width=0.4\textwidth]{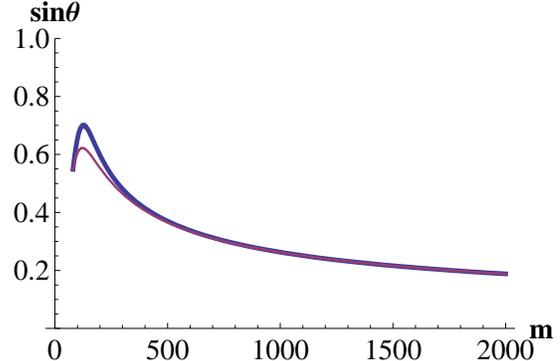}
\end{center}
\vspace{-0.25in}
\caption{The mixing angle $\sin\theta$ as a function of the mass of $N_2$ in order the relic density of $N_2$ to be equal to the dark matter density of the Universe. The thick line was derived using the improved formula of Eq.~(\ref{s}) for the $W^+$-$W^-$ channel. In addition, the annihilation channels to light fermion-antifermion pairs were included. The thin line was derived in~\cite{Kouvaris:2007iq}.}
\label{fig1}
\end{figure}

We now impose the new severe constraints of the $Ge$ detectors of CDMS on the mass of $N_2$, making a more precise calculation than in~\cite{Kouvaris:2007iq}, including the effect of the finiteness of the escape velocity from the Galaxy and the motion of the Earth. Because of the fact that the cross section of $N_2$ is fixed by demanding the relic density to be equal to the dark matter one, CDMS bounds constrain the only free left parameter, which is the mass of the particle.  $N_2$ is a Majorana particle and therefore does not scatter coherently with the nuclei. In the case of Higgs exchange, $N_2$ can scatter also coherently, but this channel is suppressed~\cite{Enqvist:1990yz}. The spin dependent cross section can be
written in convenient units pb~\cite{Kouvaris:2007iq} as
\be \sigma_{N_2} = 33.76 \times 10^{-3} \mu^2 I_s \sin^4\theta (\text{pb}), \ee where $\mu$ is the reduced mass of $N_2$ and the target nucleus measured in GeV. The constant $I_s$ is defined as
\be I_s=(C_p \langle S_p \rangle +C_n \langle S_n \rangle)^2 \frac{J+1}{J}, \ee where $J$ is the spin of the nucleus, and $\langle S_{p(n)} \rangle$ is the expectation value of the nuclear spin content due to the proton (neutron) group. The constants $C_{p(n)}$ depend on the nucleus. In the odd group model for the $Ge$ nuclei, $J=9/2$, $\langle S_p \rangle = 0$, $\langle S_n \rangle = 0.23$, and $C_n^2=0.4$ (in the nonrelativistic quark model), leading to an $I_s=0.03$. In this case the cross section can be rewritten as
\be \sigma_{N_2} = 1.01 \times 10^{-3} \mu^2
\sin^4\theta (\text{pb}). \label{cross} \ee The total rate of counts on an earth based
detector in experiments like CDMS is~\cite{Lewin:1995rx} \be
R_0=\frac{540}{A m}\left(\frac{\sigma_{N_2}}{1{\rm
 pb}}\right)\left(\frac{\rho_{dm}}{0.4{\rm GeVc}^{-2}{\rm
 cm}^{-3}}\right)\left(\frac{\upsilon_0}{230{\rm kms}^{-1}}\right),
\ee where $A$ is the atomic mass number of the nucleus of the detector,
$\rho_{dm}$ is the local dark matter density and $\upsilon_0$ is
the average velocity of the WIMP. The total rate is given in terms
of $\text{kg}^{-1}\text{days}^{-1}$ which means that for a given
detector of
 mass $x$ and of exposure time $y$, the total rate must be multiplied
 by $xy$. The current exposure of the $Ge$ detectors in CDMS is 121 kg.days. The differential rate
 with respect to the recoil energy $T$, having taken into account the motion of the Earth (through the velocity $v_E$) and the escape velocity $v_{esc}$ from the Galaxy, is~\cite{Lewin:1995rx} \be \frac{dR(v_E,v_{esc})}{dT}=\frac{1}{k}\left (\frac{dR(v_E,\infty)}{dT}-\frac{R_0}{E_0r}e^{-v_{esc}^2/v_0^2} \right ), \ee where \be \frac{dR(v_E,\infty)}{dT}=\frac{R_0}{E_0r}\frac{(\pi)^{1/2}}{4}\frac{v_0}{v_E} [\text{erf}(z_1)- \text{erf}(z_2)]. \ee $E_0=mv_0^2/2$ is the kinetic energy of the WIMP, $r=4mM/(M+m)^2$, $M$ is the mass of the $Ge$ nucleus, $z_1=(v_{min}+v_E)/v_0$, $z_2=(v_{min}-v_E)/v_0$, where $v_{min}=(T/(E_0r))^{1/2}v_0$. The constant $k$ is
 \be k = \text{erf} \left (\frac{v_{esc}}{v_0} \right )-\frac{2}{\pi^{1/2}}\frac{v_{esc}}{v_0}e^{-v_{esc}^2/v_0^2}. \ee We have chosen the following values $v_0=230~\text{km/s}$, $v_E=244~\text{km/s}$, and $v_{esc}=600~\text{km/s}$. The number of actual counts detected is given by \be \text{counts}=\int_{E_1}^{E_2}\epsilon \frac{dR(v_E,v_{esc})}{dT} dT, \ee
 where $E_1=10$ keV and $E_2=100$ keV are the lowest and highest recoil energies measured by the detectors, and $\epsilon= 121$ kg.days is the exposure. We have assumed that the efficiency of the detectors over the whole range of $T$ is 1. We should stress that the $90\%$ confidence level for the confirmed detection of one count is 2.3. In Fig.~2 we have plotted the required exposure in kg.days for the detection of one confirmed count of $N_2$ as a function of its mass. As it is evident from the figure, the new CDMS constraints currently exclude $N_2$ as a dark matter candidate probably only in a small window between 116 and 126 GeV for $\rho_{dm}=0.3$ GeV$/\text{cm}^3$, or between 95 and 156 GeV for $\rho_{dm}=0.4$ GeV$/\text{cm}^3$. As it is evident from the figure, for $m$ larger or smaller than this window, the required exposure increases very rapidly and the current CDMS constraints cannot exclude $N_2$. The reason for the increase of the exposure for larger masses is twofold. On one hand the annihilation cross section increases as a function of $m$ and therefore  the mixing angle $\sin\theta$ gets suppressed in order to give the right relic density. On the other hand, larger $m$ means smaller number density for the incoming $N_2$ particles. For masses smaller than the ``window'', the reason is the tremendous increase of the annihilation cross section as we approach the resonant value of $M_Z/2$. Again, the mixing angle $\sin\theta$ is suppressed in order for $N_2$ to account for the full dark matter abundance.

\begin{figure}[t]
\begin{center}
\includegraphics[width=0.4\textwidth]{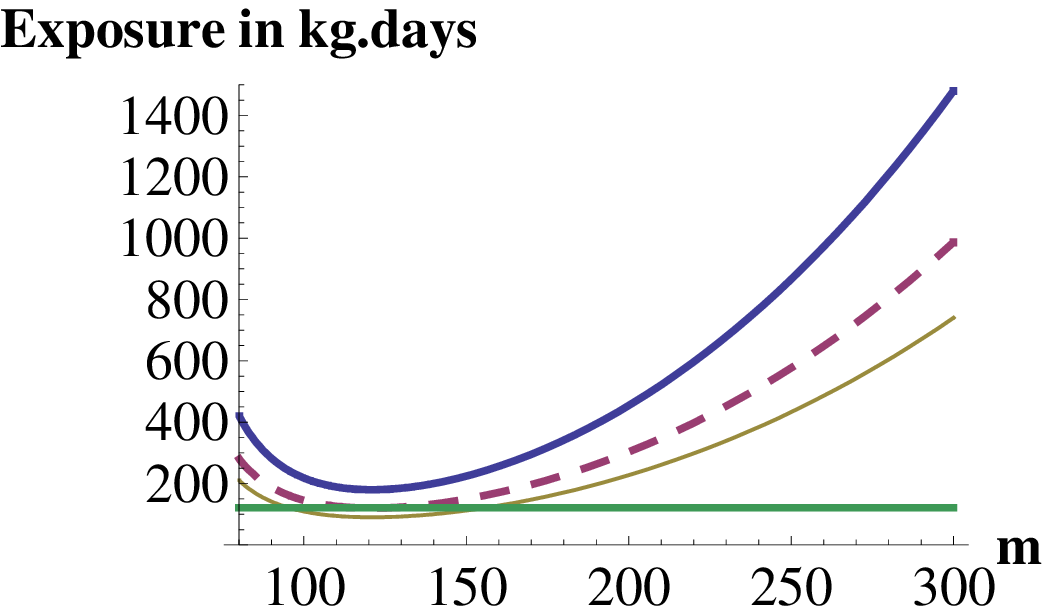}
\end{center}
\vspace{-0.25in}
\caption{The exposure of the $Ge$ detectors of CDMS required for a confirmed count of $N_2$ as a function of the mass of $N_2$. The thick, dashed and thin curves, represent the required exposure for $\rho_{dm}=0.2$, 0.3, and 0.4 GeV$/\text{cm}^3$. The horizontal line represents the current exposure in CDMS of 121 kg.days. For $\rho_{dm}=0.3$, and 0.4 GeV$/\text{cm}^3$, a small range of masses for $N_2$ has already been excluded. It is the range where the curves are below the horizontal 121 kg.days line.}
\label{fig2}
\end{figure}

 As we mentioned in the introduction, the hypercharge can be chosen such that $\nu'$ and not $D$ is the electrically neutral particle. This case was studied in~\cite{Kainulainen:2006wq}. In this scenario there is no seesaw mechanism and therefore no mixing angle $\theta$, that can allow an adjustment of the annihilation cross section in such a way that the relic density of $\nu'$ equals the dark matter density of the Universe. The relic density of $\nu'$ can be of the order of $\rho_{dm}$, if the Universe has a nonstandard expansion history including an early dominance by a rolling scalar field as predicted by many models for dynamical dark energy, like quintessence~\cite{Joyce:1996cp}. The spin dependent cross section of $\nu'$ scattering off nuclei targets is given by Eq.~(\ref{cross}) with $\sin\theta=1$. We performed the same analysis as for $N_2$. The CDMS constraints put a lower bound on the mass of $\nu'$. As it was expected, the exclusion region is larger in this case due to the fact that the cross section is larger. Our calculation shows that CDMS excludes $\nu'$ as a primary dark matter particle for a mass below 490, 760, and 1030 GeV if $\rho_{dm}=0.2$, 0.3 and 0.4 GeV$/\text{cm}^3$ respectively.

\section{The case of $\bf{He \bar{U}\bar{U}}$ or $\bf{He \zeta}$}

The minimal walking technicolor model can provide another nontrivial dark matter candidate having different properties from $DG$ or $DD$ that mentioned above. In the previous scenarios, $DD$, $DG$ or $\nu'$ behave like WIMPs with the form of Nambu-Goldstone bosons ($DD$), or of neutralino type ($DG$). However there is another interesting possibility where a stable negatively doubly charged techniparticle binds via electric Coulomb forces with a positively charged $^4He^{+2}$, forming a neutral bound state (techni-O-helium) that behaves more like a strong interacting massive particle SIMP, rather than a WIMP~\cite{Khlopov:2007ic,Khlopov:2008ty}. The scenario of having bound states between stable negatively charged particles with $He$ has been implemented in the context of different models like ones with new heavy lepton and quark families~\cite{N,Legonkov,I,lom,KPS06,Khlopov:2006dk}, tera-helium atoms~\cite{Glashow,Fargion:2005xz}, and models based on noncommutative geometry~\cite{FKS,Khlopov:2006uv}. This dark matter scenario does not require technicolor per se, however the minimal technicolor model has several advantages. The first one is that the stable charged particles emerge in a natural way, since as we already mentioned technicolor number can be protected. The second advantage is the fact that we can actually calculate the abundance of these particles since they couple to the electroweak sector and therefore they can be in equilibrium with the plasma. Finally, the strong coupling nature of the theory works effectively on the depletion of unwanted $+2$ charged techniparticles, that create problems of anomalous isotopes~~\cite{Khlopov:2007ic}.

Several issues regarding this realization of the model, as for example the effect on Big Bang Nucleosynthesis, production of anomalous isotopes, gravitational instability, possible components in cosmic rays and constraints from direct dark matter search experiments have been addressed in~\cite{Khlopov:2007ic,Khlopov:2008ty,Khlopov:2008ki}. In these studies, it was argued that techni-O-helium does not produce enough anomalous isotopes to be excluded by the current experiments, and it does not change the Standard Big Bang Nucleosynthesis picture in a significant way. In addition, it is not excluded by the data of the XQC rocket experiment. As for underground dark matter search experiments like CDMS, techni-O-helium due to fast deceleration, once being underground because of the large cross section of the helium part, it reaches the detectors with very small velocity that practically cannot transfer to the nuclei a recoil energy above the threshold of the detectors, and therefore cannot trigger an event.

\begin{figure}[t]
\begin{center}
\includegraphics[width=0.4\textwidth]{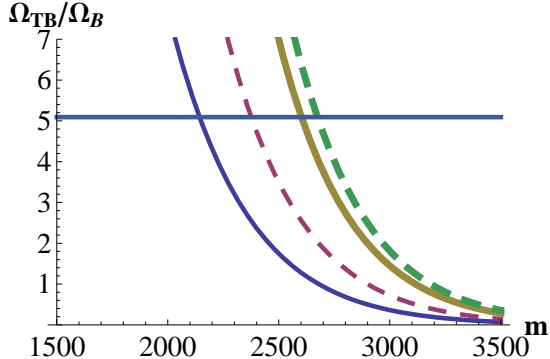}
\end{center}
\vspace{-0.25in}
\caption{$\Omega_{TB}/\Omega_B$ as a function of the mass of $\bar{U}\bar{U}$ for negligible $L'/B$ and $L/B$ equal to $-2$ (thin line), $-1$ (thin dashed line), 1 (thick line), 2 (dashed thick line). The horizontal line represents the dark matter density over the baryon density of the Universe. For the case of $He\zeta$, the plot would correspond to values of $L/B$ $-3.5$, $-4$, $-5$, and $-5.5$ respectively, if $m$ now represents the mass of $\zeta$ and $TB/B$ is negligible.}
\label{fig3}
\end{figure}

In~\cite{Khlopov:2007ic,Khlopov:2008ty}, the calculation of the abundance of techni-O-helium was done having assumed a second order phase transition for the electroweak symmetry breaking. We had also assumed that the mass splitting between $UU$, $UD$, and $DD$ was not large, unlike the situation in~\cite{Gudnason:2006yj}. The way we were able to calculate the abundance was to have an initial technibaryon-antitechnibaryon asymmetry and thermal equilibrium between the different Standard Model particles and techniparticles through weak and sphaleron processes. In this paper we address the problem of a first order phase transition. This is interesting specifically since electroweak baryogenesis scenarios require a strong first order phase transition for the electroweak symmetry breaking. The calculation was done along the lines of~\cite{Barr:1990ca,Harvey:1990qw}. The sphaleron
process~\cite{Gudnason:2006yj} involves all the three families of quarks and leptons as well as the techniquark family of $U$ and $D$ and the new lepton family. As long as the rate of the sphaleron process is larger than the expansion rate of the Universe $H$, sphalerons keep in thermal equilibrium Standard Model particles and techniparticles, and the following relation among chemical potentials is valid \be 9\mu_{u_L}+\frac{3}{2}\mu_{UU}+\mu+8\mu_W +\mu_{\nu'}=0, \ee
where $\mu_{u_L}$, $\mu_{UU}$, $\mu_W$, and $\mu_{\nu'}$ denote the chemical potentials of the left-handed up quark, $UU$, $W^-$, and $\nu'$ respectively, while $\mu$ denotes the sum of the chemical potentials of the left-handed neutrinos. We impose in our calculation overall electric charge neutrality. For a first order phase transition, the sphaleron processes freeze out before the electroweak symmetry breaking~\cite{Harvey:1990qw}, and therefore we should impose the condition $Q_3=0$, where $Q_3$ is the weak isospin charge. This condition sets $\mu_W=0$. If we assume that the isospin mass splitting among the technibaryons and the isospin splitting between the new leptons are not large, we can derive the ratio of technibaryon number $TB$ over the baryon number $B$ as \be \frac{TB}{B}=-\sigma_{UU} \left ( \frac{L}{3B}+1+\frac{L'}{3B\sigma_{\zeta}} \right ), \label{tb} \ee where $\sigma$'s are statistical factors defined in~\cite{Khlopov:2007ic}. $TB/B$ is derived as a function of the ratio of lepton over baryon number $L/B$, and new lepton family over baryon number $L'/B$. Within our approximation this result is identical to the results of second order phase transition presented in~\cite{Khlopov:2007ic,Khlopov:2008ty}, unlike the study in~\cite{Gudnason:2006yj} where $TB/B$ was different  for the two different phase transitions. This discrepancy is due to the different approximation of the mass splitting among technibaryons and between $\nu'$ and $\zeta$. The fact that within our approximation the ratio $TB/B$ is the same regardless the order of phase transition, means that the results presented in~\cite{Khlopov:2007ic,Khlopov:2008ty} hold also for a first order phase transition. We take the freeze out temperature for the sphalerons to be 250 GeV. The techni-O-helium atom can account for the whole dark matter density for a mass of the order of TeV in case we have either $He\bar{U}\bar{U}$ or $He\zeta$. In the case where both $He\bar{U}\bar{U}$ or $He\zeta$ are stable and present, no constraint is imposed on the mass, apart from the fact that the mass cannot be lower than roughly 1 TeV, since in this case it would absorb too much free $He$ and would be in disagreement with Standard Big Bang Nucleosynthesis. In Fig.~3 we plot $\Omega_{TB}/\Omega_B$ for a sphaleron freeze out temperature of 250 GeV, as a function of the mass of $\bar{U}\bar{U}$ $m$ for several different values of $L/B$ (if we assume that $L'/B$ is negligible). In order to have abundance of $\bar{U}\bar{U}$ and not $UU$, the term inside the parenthesis of Eq.~(\ref{tb}) has to be positive, putting a lower bound of $L/B>-3$ (for small $L'/B$). For typical values of $L/B$, a techni-O-helium of mass around 2 to 2.5 TeV gives the relic density of dark matter. In the case of $He\zeta$, because we want abundance of $\zeta$ and not $\bar{\zeta}$, there is an upper bound $L/B<-3$ (if now $TB/B$ is negligible). $He\zeta$ gives the right relic density for the same range of masses as $He\bar{U}\bar{U}$. The mixed scenario where both $He \bar{U}\bar{U}$ and $He\zeta$ are present~\cite{Khlopov:2007ic,Khlopov:2008ty}, gives a ratio $L/B$ close (but smaller) to $-3$ for all realistic values of the techni-O-helium masses.

We now turn to the interesting possibility of explaining the discrepancy between the DAMA and CDMS results with techni-O-helium dark matter~\cite{Khlopov:2008ty,Khlopov:2008ki}. As it was argued in these papers, techni-O-helium has too small velocity by the time it reaches the underground detectors that the recoil energy is not sufficient to trigger the detectors of CDMS. However inelastic processes of the form \be He\zeta + N \rightarrow N\zeta + He, \label{reac} \ee can take place. It is understood that a similar process can be with $\bar{U}\bar{U}$ instead of $\zeta$. $N$ is a nucleus that binds with $\zeta$ or $\bar{U}\bar{U}$ via Coulomb forces expelling $He$ from the techni-O-helium losing two electrons. If $N\bar{U}\bar{U}$ or $N\zeta$ is formed in an excited state, it can emit photons as it goes to the ground state triggering the DAMA detectors that  function on ionization. In fact it was argued in~\cite{Khlopov:2008ty,Khlopov:2008ki}, that radiation of a few keV photons can be achieved by having a transition from the $1s$ electron state of $N$ to the $1s$ electron state of $N\zeta$ (that has 2 electrons less). The binding energy of techni-O-helium is 1.6 MeV. The binding energy of $N\zeta$ and/or $N\bar{U}\bar{U}$ should be several MeV. If we use a Bohr atom description, the binding energy should be $1.6A/4$ MeV, where $A$ is the atomic number of $N$. However, the Bohr orbit is too small and this formula is inaccurate. A more detailed analysis using a harmonic oscillator potential along the lines of~\cite{Cahn:1980ss} suggests that the binding energy will be several MeV, depending on the type of $N$. We shall give an estimate for the binding energy $E$ below. If $\epsilon$ is the excitation energy from the ground state, then the energy balance of Eq.~(\ref{reac}) is \be -1.6~\text{MeV}= -E+\epsilon +E_{N\zeta}+E_{He}, \ee where $E_{N\zeta}$, and $E_{He}$ are the kinetic energies of $N\zeta$, and $He$ respectively. We omitted the initial kinetic energy of techni-O-helium and the binding energy of the two electrons of $N$ which are expelled from $N$, since they are small compared to the other energies of the problem. If we apply momentum and energy conservation for the products of the reaction~(\ref{reac}), we get \be E-1.6 -\epsilon = E_{N\zeta}\left ( 1+ \frac{m_{N\zeta}}{m_{He}}\right ). \ee Since $\zeta$ or equivalently $\bar{U}\bar{U}$ (as well as $N$ via nuclear and atomic forces) can interact coherently with the nuclei of the detectors, there is an upper bound on $E_{N\zeta}$ (or $E_{N\bar{U}\bar{U}}$), because otherwise every time there is an inelastic process of the Eq.~(\ref{reac}), $N\zeta$ in principle could pick up enough velocity to scatter off the nuclei of the detectors triggering a signal. This would mean that a signal should be found both in CDMS and DAMA. We should mention that CDMS considers $He$ signals as background. In addition CDMS veto multiple scattering events, as neutron background. However, the cross section of $N\zeta$ is so strong that it is highly unlike to have multiple detectable scattering events. A simple estimate suggests that the mean free path is so small, that although multiple scattering does occur, it cannot be identified as such by the resolution of the detectors. This suggests that in order to trigger the DAMA detectors but not the CDMS ones, the recoil energy $T=E_{N\zeta}r$ of the product of the inelastic collision $N\zeta$, should be smaller than the threshold of 10 keV of CDMS. Then the condition reads \be 4 \times 10^3\frac{E-1.6-\epsilon}{1+\frac{m_{N\zeta}}{m_{He}}}\frac{m_{N\zeta}m_T}{(m_{N\zeta}+m_T)^2}<10, \label{con1}\ee
where $E$ and $\epsilon$ are measured in MeV and $m_T$ is the mass of the target nucleus. Since $m_{N\zeta}>>m_T$ and  $m_{N\zeta}>>m_{He}$, we can simplify the condition as \be 4 \times 10^3(E-1.6-\epsilon) \frac{m_{He}}{m_{N\zeta}}\frac{m_T}{m_{N\zeta}}<10. \label{con2} \ee If we know the exact value of $E$, this formula constrains the mass of $\zeta$ (or equivalently of $UU$), or vice versa if we know $m_{\zeta}$ we can put an upper bound on $E$. Using Eq.~(\ref{con2}), for a typical value of $m_{\zeta}=2.5$ TeV, and a target nucleus $Ge$ ($N$ being also $Ge$), $E$ should satisfy $E<70$ MeV. The binding energy of $Ge\zeta$ can be estimated along the lines of~\cite{Cahn:1980ss} using an harmonic oscillator potential. $E$ is
 \be E=\frac{3}{2}\frac{Z_{N}Z_{\zeta}\alpha}{r}\left (1-\frac{1}{m_N r} \right ), \ee where $Z_N$, $Z_{\zeta}$ are the charges of $N$ and $\zeta$ respectively, $m_N$ is the mass of $N$, $\alpha$ is the fine structure constant, and $r=1.2 A^{1/3}/200~\text{MeV}^{-1}$. For the case of $Ge$ nucleus, the binding energy is 28 MeV. Using this estimate for $E$, and Eqs.~(\ref{con1}) or~(\ref{con2}), we can put a lower bound on the mass of $\zeta$ (or $UU$), such that techni-O-helium can trigger the detectors of DAMA while not the ones of CDMS. This lower bound is $\sim$ 1.5 TeV.

\section{Discussion}

In this paper we reviewed the dark matter candidates emerging from the minimal walking technicolor model. One of the candidates, i.e. $DD$ has already been excluded by the dark matter search experiments. For the case of $DG$, a bound state between $D$ and technigluons, we made a more refined analysis of the calculation of the relic density and we imposed the new CDMS constraints, taking into account the motion of the Earth and the escape velocity of the Galaxy. For this candidate, only a tiny window around 120 GeV can be excluded with 90$\%$ confidence level.

The other candidate we investigated is techni-O-helium, a bound state between either $\bar{U}\bar{U}$ or $\zeta$ with $^4He^{+2}$. We calculated the relic density in the case of a strong first order phase transition and we got the same results as for a second order calculated elsewhere. This scenario can be constrained by the ratio $L/B$, which could probably verify or rule it out once a better knowledge of the ratio will be possible. On the other hand, this scenario can explain in a natural way the discrepancy between the CDMS and DAMA experiments. We derived an inequality that the mass of the $\bar{U}\bar{U}$ or $\zeta$ and the binding energy between this particle and a nucleus from the detector should satisfy in order to have no signal in CDMS, but a positive signal in DAMA. We gave an estimate of the binding energy and we derived a lower bound for the mass of $\bar{U}\bar{U}$ or $\zeta$ at roughly 1.5 TeV.

\section{Acknowledgments}

I am very glad to thank Professor M. Yu. Khlopov for very useful discussions and for reading the manuscript. This work is supported by the Marie Curie Fellowship under contract MEIF-CT-2006-039211.


\end{document}